\documentstyle[prl,twocolumn,aps,epsf]{revtex}

\begin{document}
\wideabs{
\title{Bose-Einstein Condensation in a Surface Micro Trap}
\author{H. Ott, J. Fortagh, G. Schlotterbeck, A. Grossmann, and C. Zimmermann }
\address{
Physikalisches Institut der Universit\"at T\"ubingen\\
Auf der Morgenstelle 14, 72076 T\"ubingen, Germany
}
\date{6 July 2001}

\maketitle
\begin{abstract}

\noindent Bose-Einstein condensation has been achieved in a
magnetic surface micro trap with $4\times 10^5$ $^{87}$Rb atoms.
The strongly anisotropic trapping potential is generated by a
microstructure which consists of microfabricated linear copper
conductors at a width ranging from 3 $\mu$m to 30 $\mu$m. After
loading a high number of atoms from a pulsed thermal source
directly into a magneto-optical trap (MOT) the magnetically stored
atoms are transferred into the micro trap by adiabatic
transformation of the trapping potential. The complete
\begin{it}{in vacuo}\end{it} trap design is compatible with
ultrahigh vacuum below $2\times 10^{-11}$ mbar.
\end{abstract}
 }
\narrowtext

Trapped ultracold atoms are fascinating model systems for studying
quantum statistical many particle phenomena. Confined in optical
or magnetic trapping potentials, the atomic gas reaches quantum
degeneracy at ultra low temperatures ($< 1\mu$K) and very small
densities ($\sim 10^{14}$ cm$^{-3}$). In this regime, the
interaction between the atoms is still weak and the system is
accessible to precise theoretical description\cite{Dal}. One of
the most intriguing properties of such ultracold atomic ensembles
is the formation of macroscopic matter waves with extraordinary
large coherence lengths. For single thermal atoms the coherence
length is determined by the thermal de-Broglie wavelength and is
thus limited to the micrometer range, even for temperatures as
small as 1 $\mu$K. In contrary, a Bose-Einstein condensate may
show coherence effects over a much larger distance\cite{And,Blo}.
It is, therefore, an exciting vision to combine degenerate quantum
gases with magnetic micropotentials\cite{Wei} which may allow for
coherent atom optics on the surface of a microstructured "atom
chip"\cite{Fol}. Yet, it is an experimentally open question, how a
degenerate Bose gas behaves in a waveguide structure where it
acquires a quasi one dimensional character\cite{Pet,Dru}.

All recent experiments using microfabricated structures have been
carried out with thermal, non degenerate atomic ensembles and
therefore small coherence lengths\cite{Rei,Mue,Dek,Key,Cas,For3}.
Experiments with Bose-Einstein condensates are so far restricted
to straightforward harmonic trapping potentials with relatively
small anisotropy\cite{Mew,Mya1,Ess}. These trapping potentials are
generated by large scale current coils and can hardly be
structured on a scale much smaller then the dimensions of the
coils.  Some flexibility has been achieved by using optical
potentials\cite{Sta,Bar}, however, this approach does not allow
for long and strongly confining waveguides due to diffraction of
the employed laser light.

In our experiment we have, for the first time, generated a
Bose-Einstein condensate in a microstructured magnetic surface
trap. Our setup allows for investigating coherence phenomena of
degenerate quantum gases in extremely anisotropic magnetic
waveguides. A long lifetime of the trapped atoms and a large
number of atoms in the condensate have been achieved with a fully
\begin{it}{in vacuo}\end{it} trap design.  We optically precool
the atoms in a magneto-optical trap (MOT) which is directly loaded
from a pulsed thermal source. This fast and easy technique avoids
the standard sophisticated double MOT system \cite{Mya2}. The
surface micro trap is loaded by adiabatic transformation of the
initially shallow magnetic trapping potential into the geometry of
the micro trap.

\begin{figure}
   \begin{center}
   \parbox{8cm}{\epsfxsize 8cm\epsfbox{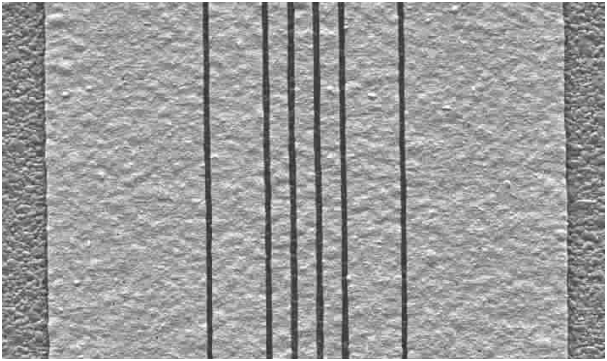} }
   \end{center}
   \caption{The Scanning Electron Microscope (SEM) image of the
microstructure shows a part of the 25 mm long Al$_2$O$_3$
substrate with seven parallel electroplated copper conductors. The
widths of the conductors are 30, 11, 3, 3, 3, 11, and 30 $\mu$m
each. The height of the copper layer is 2.5 $\mu$m.}
   \label{microstructure}
\end{figure}

The central element in our experiment is the microstructure. It
consists of seven 25 mm long parallel copper conductors at a width
of 3 $\mu$m, 11 $\mu$m and 30 $\mu$m which are electroplated on an
Al$_2$O$_3$ substrate (Fig.\ \ref{microstructure}). The total
width of the microstructure is 100 $\mu$m. Test experiments with a
set of identical microstructures revealed breakthrough currents of
1.5 A, 0.8 A and 0.4 A for the 30 $\mu$m, 11 $\mu$m and 3 $\mu$m
conductors. This corresponds to maximum current densities of
$2.0\times 10^6$, $2.9\times 10^6$ and $5.3\times 10^6$
$\mbox{A}/\mbox{cm}^2$, respectively. The current in each
conductor can be controlled individually. The configuration of the
conductors allows us to build a variety of trapping potentials
which we discuss briefly. If the currents in all conductors are
driven in the same direction, a linear quadrupole trapping
potential is formed by the combined field of the conductors and an
additional homogeneous bias field which is oriented perpendicular
to the long axis of the microstructure. By subsequently turning
off the current in the outer conductors the field gradient is
increased while the trap depth is kept constant. An even steeper
trapping potential can be achieved by inverting the currents in
the outer conductors. Then, the bias field is formed by the outer
conductors alone and no external field is required. In this
configuration the magnetic field gradient exceeds 500,000 G/cm at
the maximum possible current densities. To avoid for Majorana spin
flips, this linear quadrupole field is superimposed with a
magnetic offset field that is oriented parallel to the direction
of the waveguide. It is generated by two extra wire loops that are
mounted directly onto the microstructure substrate, one at each
end of the waveguide. With a longitudinal offset field of 1 G, the
radial curvature of the magnetic field modulus amounts to
$2.5\times 10^{11}$G/cm$^{2}$. For $^{87}$Rb atoms trapped in the
$|F=2, m_F=2>$ ground state this corresponds to a radial
oscillation frequency of $\omega_r$ = 2$\pi\times 600,000$ Hz.

The unusual length of the microstructure of 25 mm provides a large
flexibility for investigating quantum gas dynamics in long
waveguides.  A set of parallel waveguides can be formed by a
suitable choice of currents in the inner conductors of the
microstructure\cite{Hin}. The two guides may be merged and
separated by varying the strength of the longitudinal offset field
along the guides, thus allowing for the realization of on-chip
interferometers. In the experiment described in this letter we
used the microstructure in the most basic mode of operation: all
conductors are driven in the same direction and we apply an
external bias field. The radial and axial trap frequency are
adjusted to $\omega_r$ = 2$\pi\times 840$ Hz and $\omega_a$ =
2$\pi\times 14$ Hz.

\begin{figure}
   \begin{center}
   \parbox{8cm}{\epsfxsize 8cm\epsfbox{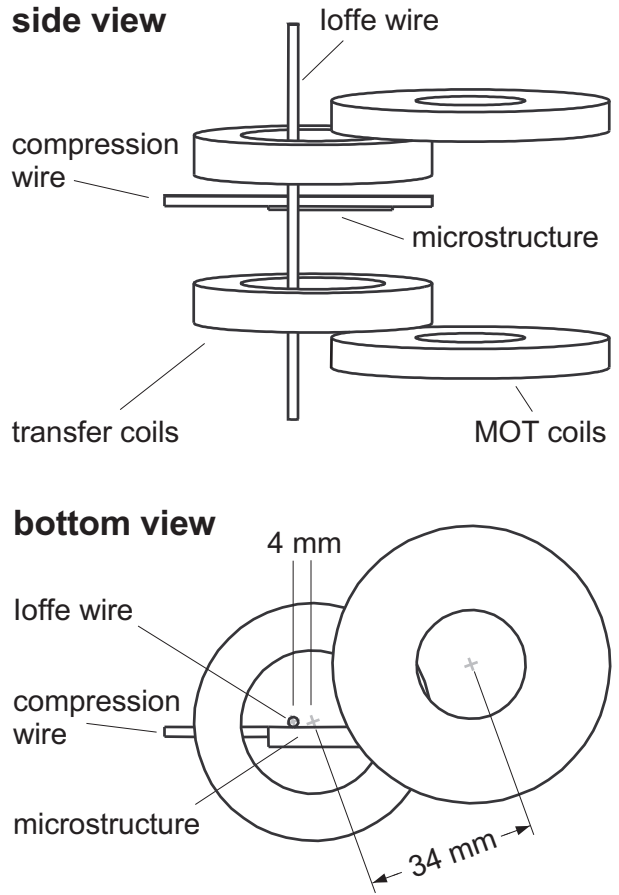} }
   \end{center}
   \caption{Trap
setup. The microstructure (Fig.\ \protect\ref{microstructure}) is
mounted upside down
 on the compression wire and the micro trap is formed below the microstructure.}
   \label{trapsetup}
\end{figure}

The adiabatic transfer of atoms into the microstructure is
achieved by means of a pair of "transfer coils". The
microstructure is mounted horizontally on a $2\times 2$ mm$^2$
copper rod ("compression wire") that is embedded in a heat sink at
the bottom of the upper transfer coil (Fig.\ \ref{trapsetup}). The
conductors on the microstructure are oriented parallel to the
compression wire. The setup is completed by a vertical copper wire
with 2 mm diameter ("Ioffe wire") that is oriented parallel to the
symmetry axis of the transfer coils but displaced by 4 mm.  At a
current of 3 A the transfer coils generate a spherical quadrupole
field with a gradient of 58 G/cm along the symmetry axis. This
field forms a relatively shallow magnetic trap where atoms can be
conveniently stored with standard techniques. By increasing the
current in the Ioffe wire, the center of the spherical quadrupole
is shifted and transformed into a Ioffe-type trapping
field\cite{For2}. At a current of 13 A in the Ioffe wire the
resulting harmonic trap potential is characterized by an axial
oscillation frequency of $2\pi\times 14$ Hz, a radial oscillation
frequency of $2\pi\times 110$ Hz and an offset field of 0.7 G. The
transfer from the Ioffe-type trap into the microstructure is
completed by changing the currents in the upper and lower coil.
This shifts the magnetic field minimum onto the surface of the
microstructure.

For the initial preparation of atoms in a magneto-optical trap we
use a second pair of coils ("MOT coils") which is separated 34 mm
from the transfer coils (Fig.\ \ref{trapsetup}). Both pairs of
coils overlap and an adiabatic transfer of magnetically trapped
atoms can be performed with the quadrupole potential minimum
moving on a straight line from the center of the MOT coils to the
center of the transfer coils\cite{Gre}. The separation of the MOT
coils and the transfer coils guarantees a good optical access to
the MOT and allows for a high flexibility in mounting different
micro trap geometries. The trap setup is placed in a UHV chamber
at a pressure below $2\times 10^{-11}$ mbar.

\begin{figure}
\begin{center}
   \parbox{8cm}{\epsfxsize 8cm\epsfbox{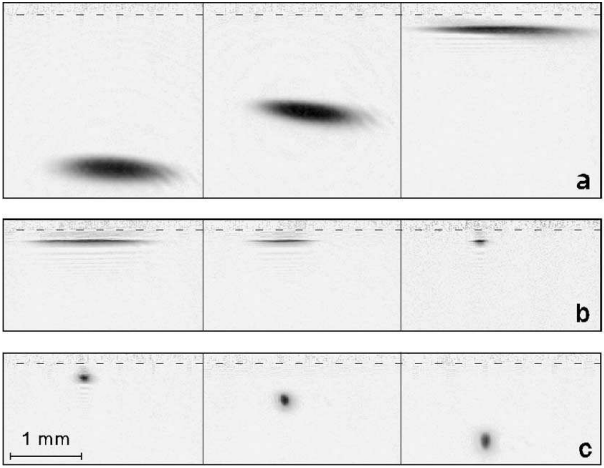} }
   \end{center}
\caption{Absorption images of the compression and the final
cooling stage. The dashed line
 indicates the surface of the microstructure. (a) Transfer and compression of the Ioffe-type
 trap into the micro trap. (b) RF cooling in the micro trap. The image on the right side
 shows the condensate in the trap. (c) Release of the condensate. The images are taken
 after 5, 10, and 15 ms time of flight.}
\label{transfer}
\end{figure}

In the present experiment we use our apparatus as follows. We load the MOT within 20 s
from a pulsed thermal source \cite{For1} ("dispenser") that is mounted at a distance of 50 mm from the
MOT. The dispenser is heated by a 12 s long current pulse of 7 A.  After the current pulse the
MOT is operated for another 8 s.  During this time the vacuum recovers and a trap lifetime of
100 s is achieved.  We collect up to $5\times 10^8$ $^{87}$Rb atoms with a six beam MOT configuration,
with 20 mm beam diameter and 20 mW laser power in each beam.  After 5 ms of polarization
gradient cooling we optically pump the atoms in the $|F=2, m_F=2>$ hyperfine ground state and
load a 70 $\mu$K cold cloud of $2\times 10^8$ atoms into the spherical quadrupole trap formed by the MOT
coils at a field gradient of 45 G/cm.  Next, the atoms are transferred into the Ioffe-type trap
within 1 s and then cooled for 20 s by radio frequency evaporation to a temperature of 5 $\mu$K.
The phase space density increases from $10^{-7}$ to $10^{-3}$, the peak density grows from
$5\times 10^{10}$ cm$^{-3}$
to $1\times 10^{12}$ $\mbox{cm}^{-3}$ and the elastic collision rate eventually reaches 60 s$^{-1}$. The precooled ensemble
is now adiabatically compressed from the large volume Ioffe-type trap into the micro surface
trap as described above (Fig.\ \ref{transfer}a). The compression is completed by inverting the current in
the compression wire and the atoms are now confined in the final trap for condensation. Best
conditions for condensation are achieved with a current of 2 A in the microstructure and
$-10$ A in the compression wire which corresponds to trap frequencies of $\omega_r =
2\pi\times 840$ Hz and $\omega_a = 2\pi\times 14$ Hz. The trap minimum is then located at a distance of 270 $\mu$m from the surface.
During transfer and compression the radio frequency is turned off.  The compression heats the
atomic cloud to 35 $\mu$K and boosts the elastic collision rate up to 600 s$^{-1}$.  By now ramping the
radio frequency from 10 MHz to 1 MHz within 7 s the phase transition occurs and Bose-Einstein condensation is achieved (Fig. \ref{transfer}b).

\begin{figure}
\begin{center}
   \parbox{8cm}{\epsfxsize 8cm\epsfbox{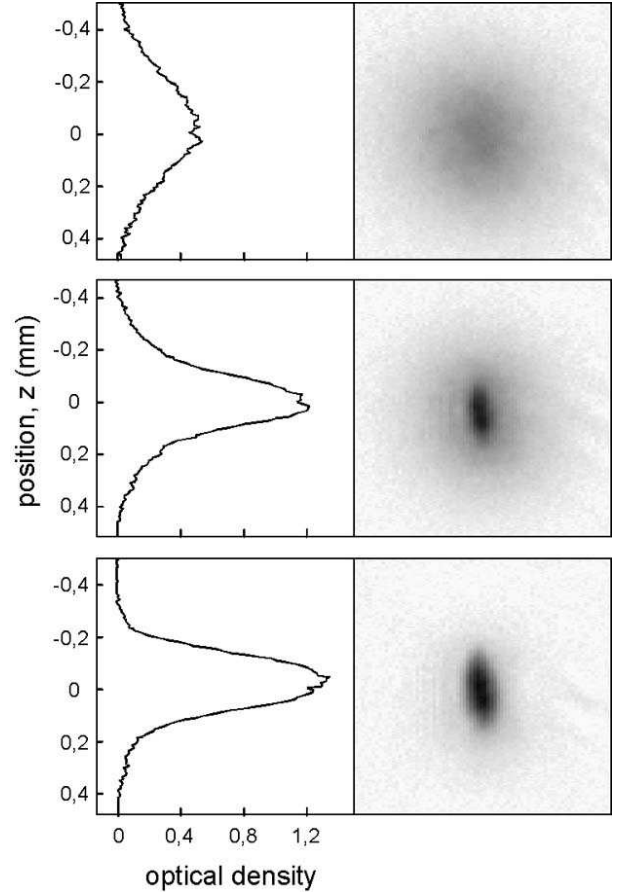} }
   \end{center}
\caption{Absorption images of the condensation. The images are
taken  after 20 ms
 time of flight at different temperatures. The left part of the picture shows a vertical
scan through the middle of the cloud. Temperature, number of
atoms, density and the chemical potential are determined by
fitting the scans. The oscillation frequencies of the trap are
$\omega_r$ = 2$\pi\times 840$ Hz and $\omega_a$ = 2$\pi\times 14$
Hz. (a) Thermal cloud: $T=970$ nK, $N=1\times 10^6$, $n_0=2 \times
10^{14}\ \mbox{cm}^{-3}$. (b) Thermal cloud and condensate
fraction: $T=730$ nK, $N=7.5\times 10^5$, $N_0=2 \times 10^{5}$,
$n_0=7\times 10^{14}\ \mbox{cm}^{-3}$, $\mu/{k_B}=285$ nK. (c)
Almost pure condensate: $T < 500$ nK, $N=5\times 10^5$, $N_0=4
\times 10^{5}$, $n_0=1\times 10^{15}\ \mbox{cm}^{-3}$,
$\mu/{k_B}=380$ nK.}\label{bec}
\end{figure}

The properties of the condensate have been determined by fitting
absorption images taken after 20 ms time of flight (Fig.\
\ref{bec}). For a pure condensate we obtain a chemical potential
$\mu/{k_B}$ = 380 nK, a density $n_0 = 1\times 10^{15}$ cm$^{-3}$
and a number of atoms $N_0 = 400,000$. We derive a critical
temperature of 900 nK for $1\times 10^6$ atoms. The lifetime of
the condensate is limited by three body collisions \cite{Bur} to
100 ms. After relaxation of the trapping potential the lifetime is
increased to 1 s.  We don't observe any heating and applying a
radio frequency shield has no effect.  Further compression of the
condensate into the micro trap leads to enhanced three body
collisions and loss of the condensate.  To optimize the number of
atoms which can be loaded into a magnetic waveguide it will be
important to study the collision properties of the condensate in
strongly anisotropic traps and quasi one dimensional situations.
Our micro trap combines the advantage of a deep magnetic trapping
potential with the possibility to vary the ratio of the
oscillations frequencies over a wide range. This offers a
promising testing ground for investigations of phase transition in
highly anisotropic traps that is in the focus of present
theoretical and experimental work\cite{Det}.

In summary, we have demonstrated that Bose-Einstein Condensation
is possible in a surface micro trap with a large number of atoms
($4\times 10^5$). The \begin{it}{in vacuo}\end{it} trap design is
compatible with UHV requirements (p $< 2\times 10^{-11}$ mbar)
where the lifetime of the atomic cloud exceeds 100 s. We have
shown that a commercially available alkali dispenser can be
effectively used as pulsed thermal source for rubidium atoms. This
represent an attractive alternative to sophisticated double-MOT or
Zeeman-slower systems.  We have introduced a novel magnetic
transfer scheme which adiabatically transforms a large volume
spherical quadrupole potential into the geometry of the small
sized parabolic micro trap with 100 \% efficiency. The evidence of
Bose-Einstein condensation in a micro trap opens a wide range of
possibilities for investigations of coherence phenomena in
waveguides and atom optical devices as well as for studies of
different regimes of the quantum degeneracy.

This work is supported in part by the Deutsche Forschungsgemeinschaft under grant no.
Zi 419/3-1.  We thank J. Schuster, M. Greiner and T. Esslinger for valuable discussions, D.
Wharam for helping with the electroplating techniques and HighFinesse for providing ultra
stable current sources.

{\it Note added in proof: In a similar experiment a group in
Munich has also achieved Bose-Einstein condensation in a magnetic
surface trap} \cite{Rei2}.

\end{document}